\documentclass[aps,prb,reprint,showpacs,superscriptaddress,twocolumn]{revtex4-1}%
\usepackage{amsfonts}
\usepackage{amsmath}
\usepackage{amssymb}
\usepackage{graphicx}

\begin{document}
\preprint{HEP/123-qed}
\title{Spin-dependent thermoelectric transport through double quantum dots}
\author{Qiang Wang}
\affiliation{Institute of Theoretical Physics and Department of Physics, Shanxi University,
Taiyuan 030006, China}
\author{Haiqing Xie}
\affiliation{Institute of Theoretical Physics and Department of Physics, Shanxi University,
Taiyuan 030006, China}
\author{Hujun Jiao}
\affiliation{Institute of Theoretical Physics and Department of Physics, Shanxi University,
Taiyuan 030006, China}
\author{Zhi-Jian Li}
\affiliation{Institute of Theoretical Physics and Department of Physics, Shanxi University,
Taiyuan 030006, China}
\author{Yi-Hang Nie}
\email{nieyh@sxu.edu.cn}
\affiliation{Institute of Theoretical Physics and Department of Physics, Shanxi University,
Taiyuan 030006, China}
\affiliation{Institute of Solid State Physics, Shanxi Datong University, Datong, 037009, China}

\begin{abstract}
We study thermoelectric transport through double quantum dots system with
spin-dependent interdot coupling and ferromagnetic electrodes by means of the
non-equilibrium Green function in the linear response regime. It is found that
the thermoelectric coefficients are strongly dependent on the splitting of
interdot coupling, the relative magnetic configurations and the spin
polarization of leads. In particular, the thermoelectric efficiency can
achieve considerable value in parallel configuration when the effective
interdot coupling and tunnel coupling between QDs and the leads for spin-down
electrons are small. Moreover, the thermoelectric efficiency increases with
the intradot Coulomb interactions increasing and can reach very high value at
an appropriate temperature. In the presence of the magnetic field, the spin
accumulation in leads strongly suppresses the thermoelectric efficiency and a
pure spin thermopower can be obtained.

\end{abstract}
\date{\today}
\maketitle

\section{Introduction}

In recent years, quantum dot (QD) systems, as a potential artificial
thermoelectric material, have attracted much attention and their
thermoelectric properties have been widely studied both experimentally
\cite{1,2,3} and theoretically\cite{4,
5,6,7,8,9,10,11,12,13,14,15,16,17,18,19,20,21,22,23,24,25,31,32,33,34,35}. Due
to deviation from the Wiedemann-Franz law and suppression of the phonon
contributions, thermoelectric efficiency in QD systems can be significantly
enhanced\cite{4,6,9}. On the other hand, many interesting thermoelectric
effects have been observed or predicted in single QD and double quantum dots
(DQDs) systems in the Coulomb blockade (CB) regime\cite{4,
5,6,7,8,9,10,11,12,13,14,15,16,17} and Kondo regime\cite{18,19,20,21,22,23,24}%
. For example, the thermopower and the thermal conductance as a function of
gate voltage present CB oscillations\cite{4}. The transient thermopower of a
single QD system can be strongly enhanced and modified under a time-dependent
gate voltage\cite{8}. In a single QD with multilevels\cite{10} or a
serial-coupled DQDs\cite{12,13}, the thermal conductance can be strongly
enhanced at high temperature due to the bipolar effects, leading to strong
suppression of thermoelectric efficiency. When interference effects are
considered in parallel coupled DQDs system, the thermoelectric efficiency can
be significantly enhanced by Fano effect and Coulomb
interaction\cite{14,15,16,33}. In particular, the strong dependence of
thermopower on temperature and gate voltage provide a sensitive way to detect
and understand the Kondo correlations\cite{21,24}.

The recent experimental observation of the spin Seebeck effect\cite{25,26,27}
inspires the research on spin effects in energy transport in QD
systems\cite{5,6,28,29,30,31,32,33,34,35} and opens a new view of thermal
manipulation of spintronic devices and the design of spin current generator.
Spin-dependent thermoelectric effects in the single QD and DQDs system
connected to two ferromagnetic leads with collinear or noncollinear magnetic
configurations is widely studied in the linear and nonlinear response
regime\cite{5,6,13,30,31,32,33}. The results indicate that the thermoelectric
coefficients are strongly dependent on the spin polarization, the relative
magnetic configurations of ferromagnetic leads and the asymmetry of dot-lead
coupling. As mentioned by J\'{o}zef Barna\'{s}\cite{6,30,33}, due to long
spin-relaxation time, the spin accumulation in leads induces spin-dependent
splitting of the chemical potential in the leads. As a result, the spin
efficiency can be larger than the charge efficiency and a pure spin
thermopower can be obtained\cite{5,30}. Up to now, most earlier theoretical
works about spin effects in heat transport mainly focused on single QD or
parallel coupled DQDs system. More recently, people start to study
thermoelectric effects in a serial-coupled DQDs system coupled to
ferromagnetic leads\cite{13}. Even so, Still much more work is needed to
understand interplay between spin effect and heat transport in serial-coupled
DQDs system, particularly in the presence of the magnetic field or the spin
accumulation in leads.

In the present work, we consider a lateral DQDs with spin-dependent interdot
coupling connected to two ferromagnetic electrodes. Spin-dependent interdot
coupling can be achieved by applying a static magnetic field on the tunneling
junction between two QDs and induces the levels spin-dependent splitting. The
tunable system properties further enrich the thermoelectronic and thermo-spin
transport properties in DQDs system. By using the nonequilibrium Green's
function technique, thermoelectric and thermo-spin coefficients are obtained.
It is found that in parallel configuration the thermoelectric efficiency can
achieve considerable values around spin-down resonance levels when the
effective interdot coupling and tunnel coupling between QDs and the leads for
spin-down electrons are small. On the other hand, in the presence of the
magnetic field, the spin accumulation in leads strongly suppresses the
thermoelectric efficiency. In particular, the thermoelectric and thermo-spin
efficiency are strongly enhanced by the intradot Coulomb interactions and can
reach very high value at an appropriate temperature. Moreover, a pure spin
thermopower can also be obtained.

The rest of this work is organized as follows. In Sec.II, we introduce the
model of lateral DQDs system with spin-dependent interdot coupling and derive
the basic analytical formulas. In Sec.III, we present the corresponding
numerical results about the influence of the related parameters on the
thermoelectric and thermospin properties in the present device. Finally, In
Sec.IV we summarize the work.

\section{Model and formalism}

We consider a system of lateral DQDs with spin-dependent interdot coupling and
two ferromagnetic leads. The magnetizations of the leads are assumed to be
collinear, and can be either in the parallel or antiparallel configuration.
The Hamiltonian describing the system can be decomposed into three parts,
$i.e.$, $H=H_{d}+H_{\alpha}+H_{T}$. The first term describes the isolated DQDs
and is given by,%
\begin{align}
H_{d}  &  =\sum_{m\sigma}\varepsilon_{m\sigma}d_{m\sigma}^{\dagger}d_{m\sigma
}+\sum_{m}Un_{m\sigma}n_{m\bar{\sigma}}\nonumber\\
&  -\sum_{\sigma}\left[  \left(  t+\delta_{\sigma}\Delta t\right)  \left(
d_{1\sigma}^{\dagger}d_{2\sigma}+d_{2\sigma}^{\dagger}d_{1\sigma}\right)
\right]  \label{1}%
\end{align}
where the operator $d_{m\sigma}^{\dagger}$ $(d_{m\sigma})$ creates
(annihilates) an electron with energy $\varepsilon_{m\sigma}$ and spin
$\sigma$ in QD $m$. We assume that the level spacing is very large so that we
can only consider a single energy level in each QD. $U$ is intradot Coulomb
interaction constant and $n_{m\sigma}=d_{m\sigma}^{\dagger}d_{m\sigma}$ is the
particle number operator. The last term sum in Eq.(1) describes the
spin-dependent tunneling between two QDs, where $\delta_{\sigma}$ is defined
as: $\delta_{\sigma}=1$ for $\sigma=\uparrow$ or $\delta_{\sigma}=-1$ for
$\sigma=\downarrow$. This spin-dependent tunneling strength can be induced by
a static magnetic field applied to the barrier region between the two QDs.
When electron passes through the potential barrier region, the electron with
spin parallel to the magnetic field have a larger tunneling probability than
one with spin antiparallel to the magnetic field due to Larmor precession of
the electron spin inside the potential barrier\cite{36,37}.

The second term $H_{\alpha}$ of Hamiltonian describes the noninteracting
electrons in the leads and is written as $H_{\alpha}=\sum_{\alpha=L,R}%
\sum_{k\sigma}\varepsilon_{k\alpha}c_{k\alpha\sigma}^{\dagger}c_{k\alpha
\sigma}$, where $c_{k\alpha\sigma}^{\dagger}(c_{k\alpha\sigma})$ creates
(annihilates) an conduction electron with energy $\varepsilon_{k\alpha}$,
momentum $k$ and spin $\sigma$ in the $\alpha$ electrode. The third term of
Hamiltonian describes the coupling between the DQDs and the electrodes and
reads $H_{T}=\sum_{k\sigma}(V_{L1\sigma}c_{kL\sigma}^{\dagger}d_{1\sigma
}+V_{R2\sigma}c_{kR\sigma}^{\dagger}d_{2\sigma}+H.c)$, where $V_{\alpha
m\sigma}$ is the hopping matrix-element between the QD $m$ and the electrode
$\alpha$, which is assumed to be independent of momentum $k$. In the following
calculation, we define the line-width matrix as $\Gamma_{nm\alpha}^{\sigma
}=V_{\alpha n\sigma}V_{\alpha m\sigma}^{\ast}\sum_{k}2\pi\delta(\omega
-\varepsilon_{k\alpha})(\alpha=L,R,$and $V_{L2\sigma}=V_{R1\sigma})$ and in a
wide-band limit, $\Gamma_{nm\sigma}^{\alpha}$ is independent of the energy.
Using the matrix representation, we have%
\begin{align}
\Gamma_{L}^{\sigma}  &  =\left(
\begin{array}
[c]{cc}%
\Gamma_{11L}^{\sigma} & \Gamma_{12L}^{\sigma}\\
\Gamma_{21L}^{\sigma} & \Gamma_{22L}^{\sigma}%
\end{array}
\right)  =\left(
\begin{array}
[c]{cc}%
\Gamma_{1L}^{\sigma} & 0\\
0 & 0
\end{array}
\right) \nonumber\\
\Gamma_{R}^{\sigma}  &  =\left(
\begin{array}
[c]{cc}%
\Gamma_{11R}^{\sigma} & \Gamma_{12R}^{\sigma}\\
\Gamma_{21R}^{\sigma} & \Gamma_{22R}^{\sigma}%
\end{array}
\right)  =\left(
\begin{array}
[c]{cc}%
0 & 0\\
0 & \Gamma_{2R}^{\sigma}%
\end{array}
\right)  \label{2}%
\end{align}
with $\Gamma_{1L}^{\sigma}=\Gamma_{11L}^{\sigma}$ and $\Gamma_{2R}^{\sigma
}=\Gamma_{22R}^{\sigma}$. When\ the spin polarization of the leads is
considered, the matrix-element may be assumed in the form: $\Gamma
_{1L}^{\sigma}=\Gamma_{L}(1+\delta_{\sigma}p_{L})$ and $\Gamma_{2R}^{\sigma
}=\Gamma_{R}(1\pm\delta_{\sigma}p_{R})$ with the sign $+$ ($-$) for parallel
(antiparallel) alignment of the magnetization in two electrodes. $p_{\alpha}$
is related to the lead's spin polarization. $\Gamma^{\alpha}$ describes the
tunneling coupling between the DQD and the $\alpha$-th electrode\ for
$p_{\alpha}=0$. In the following calculation we assume $p_{L}=p_{R}=p$, and
$\Gamma_{L}=\Gamma_{R}=\Gamma$.

By using nonequilibrium Green's function technique, the electronic and heat
current from the $\alpha$ ($L$, $R$) lead flowing into the QDs can be written
in the forms\cite{12,38,39}:%
\begin{equation}
I_{\alpha}=\sum_{\sigma}I_{\alpha\sigma}=\sum_{\sigma}\frac{e}{\hbar}\int
\frac{d\omega}{2\pi}\left[  f_{\alpha}\left(  \omega\right)  -f_{\beta}\left(
\omega\right)  \right]  T_{\sigma}\left(  \omega\right)  \label{3}%
\end{equation}%
\begin{equation}
I_{Q\alpha}=\sum_{\sigma}\frac{1}{\hbar}\int\frac{d\omega}{2\pi}\left(
\omega-\mu_{\alpha}\right)  \left[  f_{\alpha}\left(  \omega\right)
-f_{\beta}\left(  \omega\right)  \right]  T_{\sigma}\left(  \omega\right)
\label{4}%
\end{equation}
where $\beta\neq\alpha$, $f_{\alpha}=[e^{\left(  \omega-\mu_{\alpha}\right)
/k_{B}T_{\alpha}}+1]^{-1}$ is the Fermi distribution function for the $\alpha$
electrode with chemical potential $\mu_{\alpha}$, temperature $T_{\alpha}$ and
Boltzmann constant $k_{B}$. $T_{\sigma}\left(  \omega\right)  $ is the
transmission coefficient corresponding to spin channel $\sigma$ and is given
by%
\begin{equation}
T_{\sigma}\left(  \omega\right)  =Tr\left[  G_{\sigma}^{a}(\omega)\Gamma
_{R}^{\sigma}G_{\sigma}^{r}(\omega)\Gamma_{L}^{\sigma}\right]
\end{equation}
Here $G_{\sigma}^{r}(\omega)$ and $G_{\sigma}^{a}(\omega)$ are, respectively,
the retarded and advanced Green's function in the DQD in $\omega$ space, with
$G_{\sigma}^{a}(\omega)=\left[  G_{\sigma}^{r}(\omega)\right]  ^{\dagger}$.
Employing the equation of motion and adopting the Hartree-Fock truncating
approximation for higher order Green's functions, the retarded Green's
function can be obtained as from the Dyson equations%
\begin{equation}
G_{\sigma}^{r}(\omega)=\left(
\begin{array}
[c]{cc}%
G_{11\sigma}^{r}(\omega) & G_{12\sigma}^{r}(\omega)\\
G_{21\sigma}^{r}(\omega) & G_{22\sigma}^{r}(\omega)
\end{array}
\right)  =g_{\sigma}^{r}(\omega)+g_{\sigma}^{r}(\omega)\Sigma_{\sigma}%
^{r}G_{\sigma}^{r}(\omega)
\end{equation}
Where $\Sigma_{\sigma}^{r}=-i\left[  \Gamma_{R}^{\sigma}+\Gamma_{L}^{\sigma
}\right]  $ is the retarded self-energy in the wide-band limit and $g_{\sigma
}^{r}(\omega)$ is the retarded Green's function for the isolated DQD (without
coupling to electrodes) which is expressed as
\begin{equation}
g_{\sigma}^{r}(\omega)=\left[
\begin{array}
[c]{cc}%
C_{1\sigma} & t+\delta_{\sigma}\Delta t\\
t+\delta_{\sigma}\Delta t & C_{2\sigma}%
\end{array}
\right]  ^{-1} \label{7}%
\end{equation}
where $C_{m\sigma}=(\omega-\varepsilon_{m\sigma}-U)(\omega-\varepsilon
_{m\sigma})/(\omega-\varepsilon_{m\sigma}-U+Un_{m\bar{\sigma}})$ and
$n_{m\bar{\sigma}}$, the averaged occupation number of electrons with spin
$\sigma$ in the $m$-th quantum dot, can be calculated self-consistently by
using the formula $n_{m\bar{\sigma}}=\left\langle d_{m\bar{\sigma}}^{\dagger
}d_{m\bar{\sigma}}\right\rangle =-i\int\frac{d\omega}{2\pi}G_{m\bar{\sigma
},m\bar{\sigma}}^{<}(\omega)$, where $G_{\sigma}^{<}(\omega)$ is the lesser
Green function in $\omega$ space and can be obtained by the Keldysh equations
$G_{\sigma}^{<}(\omega)=G_{\sigma}^{r}(\omega)\Sigma_{\sigma}^{<}%
(\omega)G_{\sigma}^{a}(\omega)$ with the lesser self-energy $\Sigma_{\sigma
}^{<}(\omega)=i[f_{L}(\omega)\Gamma_{L}^{\sigma}+f_{R}(\omega)\Gamma
_{R}^{\sigma}]$. It should be pointed that the Hartree-Fock decoupling
approximation scheme is reasonable when the temperature during our calculation
is higher than the Kondo temperature.

In the linear response regime, the spin accumulation in electrodes can be
neglected when spin relaxation time is sufficiently short\cite{6,30,33}. In
this case, the voltage induced by temperature gradient can be treated as
independent of spin. The electric and heat currents through the system can be
written in the form%
\begin{equation}
I=e^{2}L_{0}\Delta V+\frac{eL_{1}}{T}\Delta T \label{8}%
\end{equation}%
\begin{equation}
I_{Q}=-eL_{1}\Delta V-\frac{L_{2}}{T}\Delta T \label{9}%
\end{equation}
with $L_{n}=-\sum_{\sigma}\frac{1}{\hbar}\int\frac{d\omega}{2\pi}\left(
\omega-\mu\right)  ^{n}\frac{\partial f\left(  \omega,\mu,T\right)  }%
{\partial\omega}T_{\sigma}\left(  \omega\right)  $. In the following
calculation we assume the chemical potential $\mu_{L}=\mu_{R},$ and
temperature $T_{L}=T_{R}$, thus the Fermi distribution function $f_{\alpha
}(\omega)=f(\omega)$. The electronic and thermal conductances can be
calculated by the formula $G=e^{2}L_{0}$ and $\kappa=\frac{1}{T}[L_{2}%
-\frac{L_{1}^{2}}{L_{0}}],$ respectively. The thermopower is defined as
thermoelectric voltage $\Delta V$ induced by a temperature gradient $\Delta T$
under the condition of $I=0$, \textit{i.e.}$,$ $S=\frac{\Delta V}{\Delta
T}=-\frac{1}{eT}\frac{L_{1}}{L_{0}}$. The efficiency of heat-electricity
conversion of the system is described by the dimensionless figure of merit
(FOM) $ZT=\frac{GS^{2}T}{\kappa_{ph}+\kappa}$, where $\kappa_{ph}$ denotes the
thermal conductance corresponding to the phonon contribution, which is
suppressed in such DQD-electrodes structure, and is ignored in the present paper.

If spin relaxation time of systems is sufficiently long, the spin accumulation
in electrodes should be considered\cite{6,30,33}, which leads to that the
chemical potential difference $\Delta V_{\sigma}$ between the two leads is
related to the spin channel. Thus, in this system one can define a spin bias
$\Delta V_{s}$ induced by temperature gradient via the formula $\Delta
V_{\sigma}=\Delta V+\delta_{\sigma}\Delta V_{s}$. Corresponding spin-resolved
electric and heat currents can be expressed as
\begin{equation}
I_{\sigma}=e^{2}L_{0}^{\sigma}\Delta V_{\sigma}+e\frac{L_{1}^{\sigma}}%
{T}\Delta T, \label{10}%
\end{equation}%
\begin{equation}
I_{Q}=-\sum_{\sigma}\left[  eL_{1}^{\sigma}\Delta V_{\sigma}+\frac
{L_{2}^{\sigma}}{T}\Delta T\right]  \label{11}%
\end{equation}
with $L_{n}^{\sigma}=-\frac{1}{\hbar}\int\frac{d\omega}{2\pi}\left(
\omega-\mu\right)  ^{n}\frac{\partial f_{\sigma}\left(  \omega,\mu,T\right)
}{\partial\omega}T_{\sigma}\left(  \omega\right)  $ $(n=0,1,2)$. The charge
and spin conductance are calculated as $G=e^{2}[L_{0}^{\uparrow}%
+L_{0}^{\downarrow}]$ and $G_{s}=e^{2}[L_{0}^{\uparrow}-L_{0}^{\downarrow}]$,
respectively, while thermal conductance is given by $\kappa=-\sum_{\sigma
}\frac{1}{T}[L_{2}^{\sigma}-\frac{L_{1}^{\sigma}L_{1}^{\sigma}}{L_{0}^{\sigma
}}]$, which is identical to the value without spin accumulation in such a
system. One may introduce from Eq.(10) the spin-dependent thermopower
$S_{\sigma}=\frac{\Delta V_{\sigma}}{\Delta T}=-\frac{1}{eT}\frac
{L_{1}^{\sigma}}{L_{0}^{\sigma}}$ in spin channel $\sigma$ under the condition
of current $I_{\sigma}$ vanishing\cite{6}. Then the charge and spin
thermopower are defined as $S_{c}=\frac{\Delta V}{\Delta T}=\frac{1}{2}\left[
S_{\uparrow}+S_{\downarrow}\right]  $ and $S_{s}=\frac{\Delta V_{s}}{\Delta
T}=\frac{1}{2}\left[  S_{\uparrow}-S_{\downarrow}\right]  ,$respectively.
Corresponding charge and spin figure of merit are defined as $ZT_{c}%
=\frac{GS_{c}^{2}T}{\kappa}$ and $ZT_{s}=\left\vert \frac{G_{s}S_{s}^{2}%
T}{\kappa}\right\vert $, respectively, which describe heat-to-charge-voltage
and heat-to-spin-voltage conversion efficiency of the present system when spin
accumulation is considered. $ZT_{s}$ has the form of absolute value because
the spin conductance may be negative in some regions.

\section{Numerical results and discussion}

We now numerically study spin-dependent thermoelectric effects in a DQDs
system based on the above formula, which will be presented according to three
premises: without spin accumulation in the leads and $U=0$; without spin
accumulation in the leads and $U\neq0$; with spin accumulation in the leads.
For the sake of simplicity, we assume two dots have same bare energy levels
$\varepsilon_{m\sigma}$ and are spin degenerate, i.e. $\varepsilon_{m\sigma}$
$=\varepsilon$, which can be tuned by gate voltage in experiment. The relevant
parameters are chosen as: $\mu_{L}=\mu_{R}=0$, $T_{L}=T_{R}=T$, $\Gamma=0.1$
meV, $t=2$ meV.

\subsection{\bigskip\ without spin accumulation and $U=0$}

Consider the case that the magnetizations of the leads are in parallel
configuration, there is no spin accumulation in the leads and the intradot
Coulomb interactions are ignored. The tunnel coupling between two QDs is
spin-dependent due to the external magnetic field, such a coupling DQDs can be
equivalent to a single QD with four spin-dependent levels, $E_{1\uparrow
}=\varepsilon+t+\Delta t$, $E_{2\uparrow}=\varepsilon-t-\Delta t$,
$E_{3\downarrow}=\varepsilon+t-\Delta t$, and $E_{4\downarrow}=\varepsilon
-t+\Delta t$, which corresponds to poles of the Green function $g_{\sigma}%
^{r}(\omega)$ given by Eq.(7). These levels are identified as $E_{u\uparrow}$
(u=1,2) occupied by spin-up electrons and $E_{d\downarrow}$ (d=3,4) occupied
by spin-down electrons, and satisfy $E_{1\uparrow}>E_{3\downarrow
}>E_{4\downarrow}>E_{2\uparrow}$ for chosen parameters. \

The influence of splitting $\Delta t$ of interdot coupling on the
thermoelectric transport is shown in Fig. 1. Four resonance peaks appear in
the electrical conductance spectrum when the spin-dependent level crosses the
Fermi level ($%
\mu
=0$), see Fig. 1(a). It can be clearly seen that the variety of $\Delta t$
only changes the electrical conductance's resonance positions and do not
influence its magnitude. It is found that this hold true for the thermal
conductance $\kappa$ at low temperature. However, the behavior of $\kappa$ is
more complicated at relatively high temperature (shown in Fig. 1(b)), and the
thermopower $S$ and the figure of merit $ZT$ are more sensitive to the variety
of $\Delta t$. The antisymmetry, oscillation and sign reversal of $S$ also are
obtained from Fig.1(c). As shown in Fig. 1(d), $ZT$ presents typical
double-peak structure. When $\Delta t$ increases, for the two peaks of $ZT$
located around $\epsilon=E_{d\downarrow}$, not only their magnitudes augment
but also their degree of asymmetry increases. In particular, $ZT$ exhibits a
behavior similar to crossover and can reach very high value when $\Delta
t\ $is close to $t$ (see the inset of Fig. 1(d)). This is because when $\Delta
t\ $is close to $t$, the levels $E_{3\downarrow}$ and $E_{4\downarrow}$ get
much closer to each other and the spin-down resonance channels are
significantly blockaded due to the strong reduce of the interdot tunneling for
spin-down electrons. While with $\Delta t$ increasing the side peaks located
at $\epsilon=E_{d\uparrow}$ only shift their positions with unchanged the
maximum.

Next, we study the influence of spin polarization $p$ on the thermoelectric
transport. The relevant thermoelectric quantities versus dot's level position
$\varepsilon$ for different spin polarization $p$ are presented in Fig. 2.
When $p$ increases, the electrical and thermal conductance peaks (the
thermopower peaks) located at $\epsilon=E_{u\uparrow}$ increase (decreases)
while the electrical and thermal ones (the thermopower ones) located at
$\epsilon=E_{d\downarrow}$ decrease (increases) because $\Gamma_{1L}%
^{\uparrow}$, $\Gamma_{2R}^{\uparrow}$ increases and $\Gamma_{1L}^{\downarrow
}$, $\Gamma_{2R}^{\downarrow}$ decreases. These properties of $G$, $\kappa$
and $S$ determine the behavior of $ZT$. As shown in Fig. 2(d), with increasing
$p$ the $ZT$ near $\epsilon=E_{u\uparrow}$ is suppressed while the $ZT$ near
$\epsilon=E_{d\downarrow}$ is intensively enhanced. The dependence of $ZT$ on
spin polarization in the leads and on spin-dependent levels may be useful for
designing spin-dependent thermoelectric devices. However, it should be noted
that when the spin polarization in leads is very high, for example, $p>0.95$,
the tunneling rate for spin-down electrons is close to zero leading to
significant suppression of transport through the spin-down levels and thus the
corresponding $ZT$ peaks rapidly decrease with increase of $p$, which can be
clearly seen from the inset (2) of Fig. 2(d).

When the magnetic configuration changes from parallel to antiparallel, the
electrical conductances are suppressed for four resonance channels, but the
changes of the thermal conductance $\kappa$ and the thermopower $S$ are
dependent of the resonance levels (see Fig. 3). $\kappa$ ($S$) lowers
(heightens) for the spin-up resonance channels and heightens (lowers) for the
spin-down resonance channels. As a result, $ZT$ in antiparallel configuration
presents higher peaks around the spin-up resonance channels and lower peaks
around spin-down resonance channels comparing with in parallel configuration.
For antiparallel configuration, the tunneling rates between the leads and the
QDs satisfy $\Gamma_{1L}^{\uparrow}\Gamma_{2R}^{\uparrow}=\Gamma
_{1L}^{\downarrow}\Gamma_{2R}^{\downarrow}$, which leads to the transmission
coefficient $T_{\sigma}\left(  \omega\right)  =|G_{\sigma12}^{r}(\omega
)|^{2}\Gamma_{1L}^{\sigma}\Gamma_{2R}^{\sigma}$ is spin-independent. Then the
thermoelectric quantities have the same intensities at different resonance
points. Although the electric and thermal conductance experience an decreasing
trend with increasing $p$, the thermopower and the figure of merit are
independent of $p$. Based on the discussions above, in order to obtain high
thermoelectric efficiency one should tune the resonances points to spin-down
levels and take $p\symbol{126}0.9$ in parallel configuration for fixed
temperature and splitting of interdot coupling.

\subsection{without spin accumulation and $U\neq0$ \bigskip}

Now we focus on the case taking into account the intradot Coulomb
interactions. Under the Hartree-Fock approximation, such a coupling DQDs
system can be equivalent to a single QD with eight energy levels, in which
four levels are occupied by spin-up electrons and other four are occupied by
spin-down electrons\cite{36}. To determine the eight spin-dependent effective
energy levels one need self-consistently solve the equation $\left\vert
[g_{\sigma}^{r}(\omega)]^{-1}\right\vert =0$ due to $U\neq0$. Fig. 4 presents
the relevant thermoelectric quantities versus the dots level position
$\varepsilon$ with different temperature and $U=5$. At low temperature eight
resonance peaks of $G$ and $\kappa$ can be clearly seen from Fig. 4(a) and
(b). The peaks $1$, $4$, $5$, $8$ ($2$, $3$, $6$, $7$) correspond to the
electron transport through spin-up (down) resonance channels. With the
temperature increasing the conductance peaks become wider and lower due to
broadening of the Fermi distribution function. More interestingly, at high
temperature, the thermal conductance presents five high peaks due to the
bipolar effects\cite{10,12}.

$S$ and $ZT$ experience a nonmonotonous variation trend with increasing
temperature (see Fig. 1(c), (d)). In particular, at high temperature, two high
peaks of $ZT$ appear near symmetry point $\varepsilon=-U/2$, resulting from a
suppression of $\kappa$ and an enhancement of $S$. Comparing with the case in
the absence $U$, it can be found that Coulomb interactions shift the positions
of the peaks of $ZT$ and strongly enhance $ZT$. As shown in the insert of Fig.
4(d), The maximum of $ZT$ increases with Coulomb interactions increasing and
can reach very high values above 30 at an appropriate temperature, which is
very useful for the thermoelectric energy conversion devices. It should be
mentioned that with varying the relative magnetic configurations and the spin
polarization of leads, the relevant thermoelectric quantities exhibit same
variation trend as that without Coulomb interactions except for peaks around
$\varepsilon=-U/2$, which are unaffected.

\subsection{with spin accumulation}

For the systems with long spin relaxation time, the spin accumulation in
electrodes should be considered\cite{6,30,33}. Fig. 5 presents the
thermoelectric coefficients versus $\varepsilon$ in the presence of spin
accumulation. It is found by comparing Fig.(3) with Fig.(5) that the spin
accumulation in electrodes has no influences on the electrical conductance and
the thermal conductivity (not shown) but strongly influences on the
thermopower and the figure of merit. The peaks of $S_{c}$ and $ZT_{c}$ are
much lower than the appropriate peaks in $S$ and $ZT$. These properties are
quite different from those in the absence of external magnetic field, i.e. the
splitting of interdot coupling $\Delta t=0$. Since the levels are strongly
spin-dependent in such system, the spin FOM experiences the same variation
trend as charge FOM with varying the relative magnetic configurations and the
spin polarization of leads, and the behaviors are similar to the case without
spin accumulation discussed above.

Some novel properties also appear, for instance, in the absence of Coulomb
interactions, $G_{s}=-G$ at resonance points $\epsilon=E_{d\downarrow}$ while
$G_{s}=G$ at resonance points $\epsilon=E_{d\uparrow}$ (see Fig. 5(a)). More
interestingly, for certain $\varepsilon$ (pointed by arrow and being around
$\varepsilon=\pm t$ in Fig. 5(b)), the charge thermopower vanishes while the
spin thermopower can be finite because the thermopowers for different spin
have same value but opposite sign, which means that a pure spin current
without charge current may be produced by a temperature gradient in our
system. This feature can also be obtained in the spectrum of figure of merit
in Fig. 5(c). When the Coulomb interaction is considered, near the symmetry
point $ZT_{c}$ exhibits two very high peaks while $ZT_{s}$ completely
vanishes. This is because the thermopower for different spin have the same
value and sign leading to the enhancement of charge thermopower and the
suppression of the spin thermopower. Even so, $ZT_{c}$ and $ZT_{s}$ are
strongly enhanced by the Coulomb interactions, and $ZT_{s}$ can be larger than
$ZT_{c}$ for selected dots level positions in Coulomb blockade regime. This
implies our proposed system is a good heat-to-spin-voltage converter.

\section{Conclusion}

In summary, we studied spin effects in thermoelectric transport through a
lateral DQDs system with spin dependent interdot coupling and ferromagnetic
electrodes. It is found that in parallel configuration the thermoelectric
efficiency can achieve considerable values around spin-down resonance levels
when the effective interdot coupling and tunnel coupling between QDs and the
leads for spin-down electrons are small. On the other hand, in the presence of
the magnetic field, the spin accumulation in leads strongly suppresses the
thermoelectric efficiency. The thermoelectric and thermo-spin efficiency are
strongly enhanced by the intradot Coulomb interactions and can reach very high
value at an appropriate temperature. Moreover, In such a DQDs system a pure
spin thermopower can be obtained.

\section*{Acknowledgment}

This work was supported by National Nature Science Foundation of China (Grant
Nos. 10974124 and Grant Nos. 11004124) and Shanxi Nature Science Foundation of
China (Grant No. 2009011001-1).

\newpage

\bigskip

Figure Caption:
Fig. 1 (Color online) Thermoelectric coefficients (a) $G$, (b) $\kappa$, (c) $S$, and (d) $ZT$ as a
function of dot's level position $\varepsilon$ for varied splitting $\Delta t$
in parallel configuration with $k_{B}T=0.1$ meV. $\Delta t$ is measured in
meV. Other parameters are taken as: $t=2$ meV, $U=0$ meV, $\Gamma=0.1$ meV,
and $p=0.5$. The inset in (d) depicts $ZT$ as a function of $\varepsilon$ and
$\Delta t$.

Fig. 2 (Color online)
Thermoelectric coefficients (a) $G$, (b) $\kappa$, (c) $S$, and (d) $ZT$
versus the dot's level position $\varepsilon$ for different spin polarization
$p$ in parallel configuration with $k_{B}T=0.01$ meV. $\Delta t=1$ meV, other
parameters are the same as in Fig.1. The inset (1) in (d) is an enlargement of
the region near $\varepsilon=-1$ meV. The inset (2) depicts $ZT$ as a function
of spin polarization $p$ with fixed $\varepsilon=-1.03$ meV.

Fig. 3 (Color online) Thermoelectric coefficients
(a) $G$, (b) $\kappa$, (c) $S$, and (d) $ZT$ versus the dot's level position
$\varepsilon$ in different configuration: antiparallel configuration and
parallel configuration with $p=0.5$. Other parameters are the same as in
Fig. 2. Inset: enlargement of the region near $\varepsilon=-1$ meV.

Fig. 4 (Color online)
Thermoelectric coefficients (a) $G$, (b) $\kappa$, (c) $S$, and (d) $ZT$
versus the dot's level position $\varepsilon$ for varied temperature $T$ in
parallel configuration with $U=5$ meV and $p=0.5$. The energy parameter
$k_{B}T$ is measured in meV. Other parameters are taken as in Fig. 2.

Fig. 5 (Color online) (a)
Electrical conductance $G$ and spin conductance $G_{s}$, (b) thermopower $S$
and spin thermopower $S_{s}$, (c) (d) charge figure of merit $ZT_{c}$ and spin
figure of merit $ZT_{s}$ versus $\varepsilon$ in parallel configuration with
$p=0.5$. The Coulomb interactions $U=0$ meV in (a) (b) (c) and $U=5$ meV in
(d), Other parameters are the same as in Fig. 2.

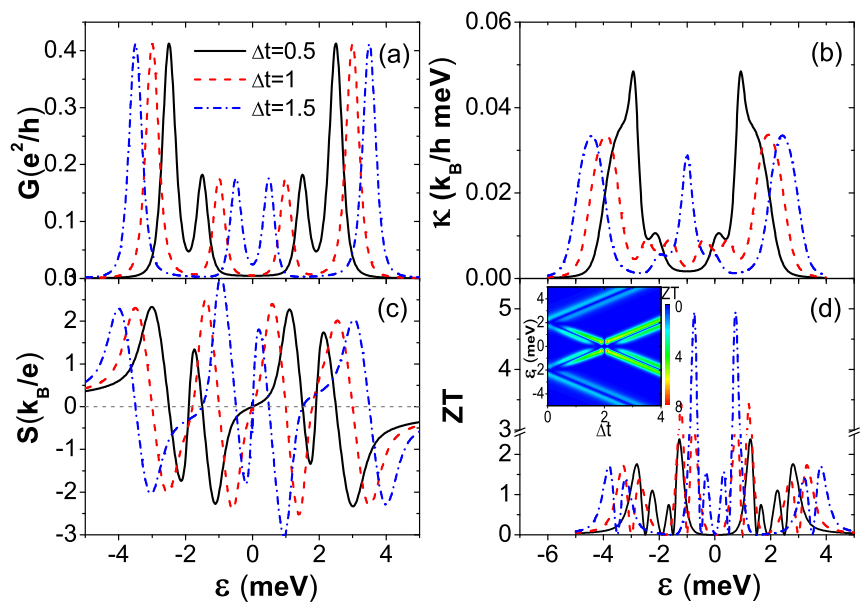
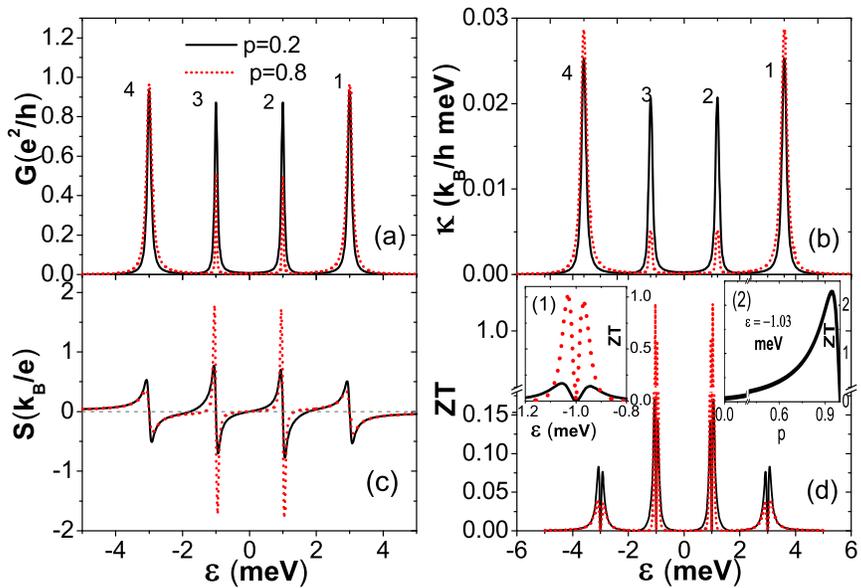
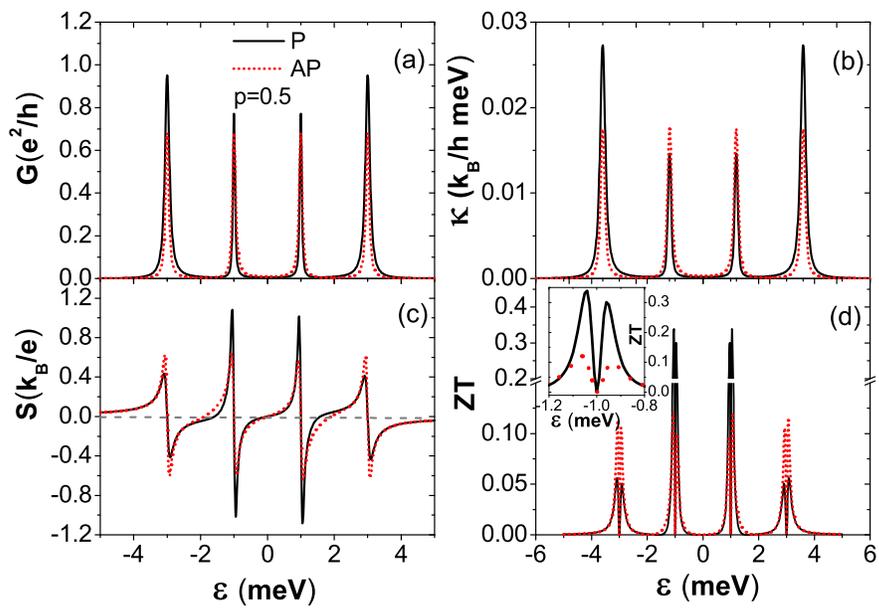
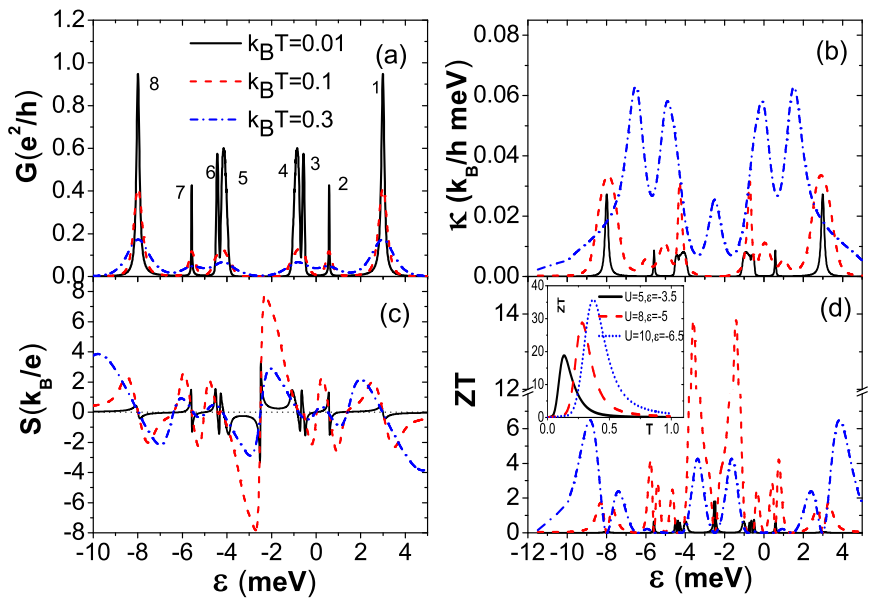
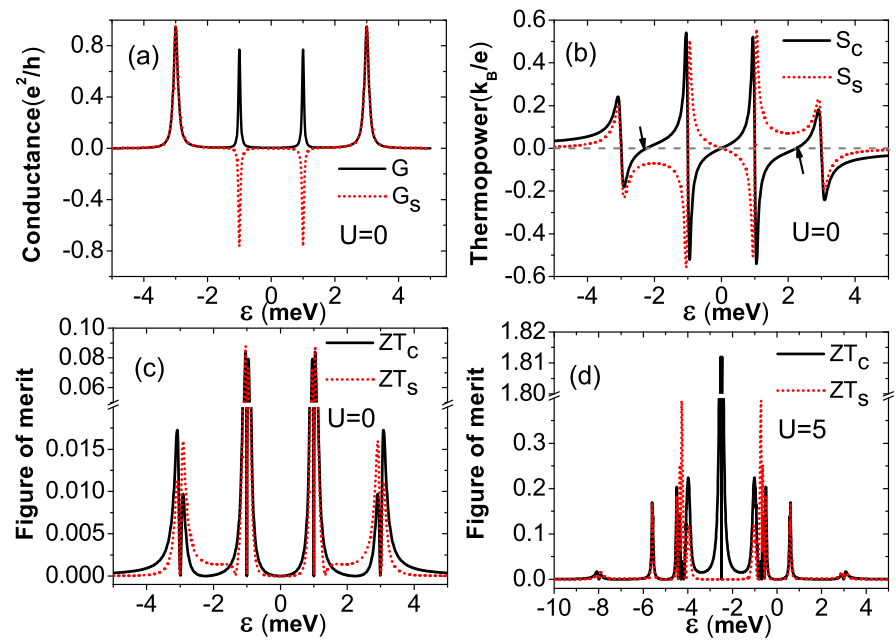
\end{document}